# Periodic ripples on thermally-annealed Graphene on Cu (110) – reconstruction or Moiré pattern?

Colm Durkan

Nanoscience Centre, University of Cambridge, 9 JJ Thomson Avenue, Cambridge CB3 0FF, United Kingdom
email: cd229@eng.cam.ac.uk

(Dated: October 9th 2017)

We have used Ultrahigh Vacuum (UHV) Scanning tunneling microscopy (STM) to investigate the effect of thermal annealing of graphene grown by chemical vapour deposition (CVD) on a Cu(110) foil. We show that the annealing appears to induce a reconstruction of the Cu surface along the [210] direction, with a period of 1.43 nm. Such reconstructions have been ascribed to the tensile strain induced in the Cu surface by differential thermal expansion of it relative to the graphene overlayer, but we show that it is in fact a Moiré pattern due to interference between the graphene and the underlying atomic lattice as evidenced by the appearance of an odd-even transition only observed due to mis-orientation of the top layer of a crystal. This highlights that the analysis of STM measurements of graphene on metal surfaces should take such interference into account and that the graphene-Cu interface is more complex than previously thought.

Graphene is being widely used in a variety of application spaces requiring the unique mechanical, electrical or optical properties that it has to offer [1]. The turning point came with the development of large-scale growth capabilities. Initial work on mechanically exfoliated graphene enabled many of the fundamental properties of interest to be probed in one-off devices [2, 3]. This was improved later via chemical processing techniques to separate graphene layers from graphite, although this led to the production of graphene oxide [4]. Nonetheless, this has proved useful for the production of graphene for composites and inks [5, 6]. The most important development was the discovery that graphene could be grown on metal substrates by the thermal decomposition of simple hydrocarbons such as ethylene [7-9]. The most commonly used substrate for growth is polycrystalline Cu foil as it tends to have the highest yield for device-grade graphene, i.e. large (> 100 µm across) graphene grains with a low number of defects, so they are capable of possessing high mobility. There have been very many reports on the nanometer and atomic-scale structure of graphene on Cu that show varying degrees of interaction between the two materials. In some cases, there is a clear Moiré pattern due to interference between the periodic electronic density of states in both graphene and Cu indicating a relatively strong interaction between the two [10], while others have reported on the ability to image the graphene lattice, indicating a very weak interaction [11]. It is of course also possible to image the underlying Cu lattice without being able to observe the graphene. Ultimately, the intrinsic interaction between graphene and Cu is expected to be rather low due to (i) the low electron density in graphene and (ii) the presence of a thin oxide on the Cu, although if this is removed immediately prior to graphene growth it appears to act as a barrier against oxidation [12]. However, due to residues from sample processing and the existence of contaminants which lead to the unintentional doping of graphene, this interaction can be a lot stronger than otherwise expected. For this reason, most experimental studies on graphene resort to two options in an attempt to mitigate against this – either encapsulate the graphene in a thin layer of oxide such as $TiO_3$ or $Al_2O_3$, or thermally anneal it under UHV conditions to desorb contaminants, and maintain the vacuum for the duration of the given experiment. This thermal annealing process, depending on the temperature, leads to multiple processes occurring simultaneously: (1) contaminants desorb from the graphene, (2) Cu oxide can be removed, (3) strain arises both due to thermal expansion of Cu and contraction of graphene and (4) Cu atoms on the surface become more mobile. It is the latter process in particular that is of interest to us in this study, as it has been proposed that this can lead to a reconstruction of the Cu surface [12]. This reconstruction is direct evidence of a very strong interaction between graphene and the surface, at least during the heating process. It is also known that during high-temperature CVD growth of graphene on Cu, the differential thermal expansion coefficients of both materials leads to a compressive strain on graphene, which in turn leads to a macroscopic reconstruction of the Cu surface via step bunching [13, 14]. In this paper, we will concentrate on what happens at the atomic scale and we will show that results such as those shown in ref [12] and ascribed to a reconstruction of the Cu can also be explained as being due to a Moiré pattern.

We obtained graphene on predominantly (110) polycrystalline Cu foil from a commercial supplier [15]. The sample was transferred to the UHVSTM (Omicron AFM/STM system) with a base pressure of $10^{-10}$ mbar, and degassed at 120$^{o}$C for 18 hours. An STM image of a large area (495 nm) as shown in Fig. 1 (a) reveals a number of characteristics of the un-annealed surface all in the same area. We can see (i) folds in the overyaler of graphene as indicated by the arrows, (ii) the cubic symmetry of the Cu(110) surface in the step morphology, as well as the atomically-resolved Cu lattice with the expected atomic spacing of 0.25 nm along the [1$\bar{1}$0] rows and 0.36 nm between the rows in the [100] direction, as indicated in area "*A*", and shown in Fig. 1(b), with an apparent atomic corrugation of 20-30 pm. In some areas, it is possible to obtain atomically-resolved images of the graphene as shown in Fig. 1(c), and the characteristic large-scale features of graphene, labelled "*G*" as shown in Fig 1(d) where we show a ripple (indicated by the solid arrow) and a grain boundary/tear (dotted arrow) in close proximity to each other. There are other areas such as "*B*" where it is not possible to resolve either the Cu or the graphene atomic structure which will almost certainly be due to the presence of a thin oxide layer on the Cu in those regions. The vast majority of the morphology that we observe is due to the underlying Cu surface. Due to the large number of steps, it is clear that the contact between the graphene and the Cu will be highly variable, leading to variations in strain in the graphene and different interaction strengths between the graphene and the Cu at different locations on the surface. We expect that this will be significant in determining which areas are most affected during any thermal annealing processes.

Having thus characterized the surface by STM, the sample was placed in a heater in the same UHV chamber as the STM, and heated to 400°C for 24 hours. The gross morphology of the surface appeared relatively unchanged apart from the fact that at a number of locations, a parallel stripe pattern had emerged. This is similar to that reported elsewhere [10, 12] and has been ascribed to a reconstruction of the Cu surface, which itself is well-known to readily form a variety of reconstructions under the influence of adsorbates [16-18]. In Fig. 2(a), we show an example of this pattern showing that these stripes are consistent over several nm. This is also clearly visible in the 3D rendering of the same area as shown in Fig. 2(b). The period is 1.43 nm, significantly less than that reported by others in the literature (5-7 nm period has been observed), but greater than that expected due to diffusion of chemisorbed O across the surface interacting with mobile Cu adatoms either forming additional rows or removing ones [18]. The apparent corrugation of the stripes is of order 50 pm, similar to that seen elsewhere. From the atomically-resolved image and by comparison with Fig 1(b), we can see that the stripes form at 55° to the [1$\bar{1}$0] raised atomic rows, so are close to one of the [210] directions (which area at 60°), where the spacing between adjacent atoms on the top surface (i.e. diagonally across the rows) is 0.443 nm. Therefore, with the undistorted lattice, there is in fact no way to form a stable reconstruction with this period in any direction. In the direction we observe, the closest corresponds to 3 periods along the [210] direction, corresponding to 1.33 nm as shown in Fig. 3, which shows a simulated view along the [210] direction for both the original and distorted Cu(110) surface. In order to achieve the observed periodicity, each surface atom would need to be displaced laterally by approximately 0.1/6 nm ~ 17pm or 4% of the interatomic distance. While this is not unreasonable and is associated with 0.1-0.3 eV of energy [19], what we appear to observe is the formation of a series of stacking faults, as is clear from Fig. 2(a), and indicated by the arrow, where the reconstruction comprises sections approximately 2.1 nm long followed by a slip of one atomic distance. The temperature at which this reconstruction is formed is rather low at 400 °C which corresponds to 58 meV of thermal energy. For comparison, the well-known herringbone reconstruction of Au(111) is formed at temperatures in excess of 580 °C, which corresponds to 73 meV of energy. We should bear in mind that the surface diffusion coefficient for any material follows an Arrhenius form, as given in Equation 1.

$$D(T) = D(0)e^{-\frac{E}{k_B T}} \qquad \text{Equation 1}$$

For the Cu(110) surface, the activation energy, $E$ = 230 meV and the pre-exponential coefficient is 8x10$^{-4}$ cm$^2$/s [20]. For comparison, the same variables for Au(111) are 21 meV and 2x10$^{-4}$ cm$^2$/s [20]. At the relevant temperatures mentioned above, the surface diffusion coefficient for Cu is an order of magnitude smaller than that of Au and coupled with the fact that the thermal energy is sufficiently larger than the activation energy for Au, the surface atoms can move relatively freely on that surface in order to form this complex reconstruction. Therefore, we would simply not expect the Cu(110) surface to form a reconstruction involving significant lateral movement of the atoms at this low temperature and in the absence of adsorbates. It is also known that graphene acts as an essentially impermeable barrier so it is unlikely that there are adsorbates on the Cu surface, and that in fact the only interaction is between it and graphene. This is only partially offset by the fact that the surface free energy of Cu(110) is 1.79 J/m$^2$ as compared to 1.51 J/m$^2$ for Au(111) [20], which means that by and large, the Cu surface is inherently more reactive than the Au one. The free energy of different basal planes will be different, and that of the (110) surface of Cu is the highest, and this is of course the reason why this is the most commonly used surface on which to grow CVD graphene as this surface catalyses the decomposition of the hydrocarbon feedstock of choice. All of this points in the same direction – from an energetic perspective, the (110) surface is less stable than other surfaces, so it will be prone to forming reconstructions under suitable conditions, but the energy provided thermally is not sufficient to overcome the activation energy to do so. We should therefore consider alternative mechanisms, for instance that the heating induces a tensile strain on the Cu due to the differential thermal expansion of it and graphene (this difference is of the order 20 ppm), which leads to a thermomechanical stress of the order 3 GPa, which is similar to the change in surface stress associated with the formation of some surface reconstructions [23]. It is well-known that this differential expansion (Cu expands upon heating whereas Graphene contracts) leads to ripple formation and strain in the graphene, although this is generally with much larger periods than we observe here [24, 25]. The resultant strain on Cu under higher-temperature annealing of graphene is also known to lead to step-bunching, so it does lead to large-scale reconstruction of the Cu surface and so should be considered as a possible trigger for the sort of features we are observing. Therefore, the possibilities are that (i) an atomic-scale surface reconstruction is indeed formed at these low temperatures due to the increased surface strain upon heating and (ii) this is not a reconstruction at all and previous attempts at analyzing this in this way have been flawed, and it is in fact simply either ripples in the graphene or a Moiré pattern due to interference between the graphene and the underlying Cu lattice. The very fact that the stripe pattern we observe is not exactly along any of the Cu atomic row directions is a clear indication that this is indeed a Moiré pattern. Although the two crystals are incommensurate, there are a number of possible Moiré patterns that can form, and as we show in Fig. 4, there is indeed one misorientation angle (5$^o$) between the graphene and Cu lattices at which a linear pattern forms roughly along the [210] direction (at the angle observed) and with a period of 1.43 nm, and which displays the same odd-even transition [26, 27] every 2.1 nm. This so-called transition which manifests itself as an apparent shift by one atomic position along the atomic rows is associated with a rotation of the top layer of a material relative to the layers underneath and has been observed many times on graphite by STM. It is only observable when the misorientation of the top layer is relatively small, as is the case here. This is illustrated in Figs. 4(b) and (c) where we show an experimental and simulated section of the graphene/Cu(110) system. The locations of the transition are indicated by arrows. This unequivocally shows that one must be careful when interpreting STM images of this complex system.

**Conclusions**

We have used UHVSTM to explore the effect of heating of the Cu(110) surface which has CVD graphene grown on top. Upon annealing for a prolonged period at a low temperature of 673 K, the surface is observed to form a linear stripe pattern. Such patterns have in the past been ascribed to a reconstruction of the surface or Moiré fringes with no clear analysis. In this article, we have shown how to determine which of these mechanisms is applicable by systematically considering the energetics and by carefully analyzing atomic-scale images, looking for characteristic features such as the odd-even transition often seen in graphite, and which is a characteristic feature of Moiré patterns.

**FIGURE CAPTIONS**

**Fig. 1**. (a) STM image of a 490 nm x 490 nm area of graphene on Cu(110) before annealing. The arrows indicate the position of ripples/folds in the overlayer of graphene. In some areas such as "*A*", the underlying Cu lattice is seen, and in others, marked "*B*", no atomic features are discernible. In most areas (indicated by "*G*", the graphene atomic lattice is visible; (b) atomic resolution on Cu(110) from region "*A*" showing the raised atomic rows along the [1$\bar{1}$0] directions; (c) atomic resolution on graphene from a region near a fold; (d) detail of area surrounding a fold (solid line) and a tear (dashed line) in graphene.

**Fig. 2.** (a) the Cu(110) surface after annealing revealing periodic features almost along the [210] direction. One of the regions where the pattern shifts by one atomic spacing is illustrated by the arrow; (b) 3D rendering of the surface, as seen along the [210] direction; (c) height cross-section perpendicular to the pattern showing the period of 1.43 nm and corrugation of around 50 pm.

**Fig. 3.** Schematic of the Cu(110) surface along the [210] direction showing (a) the bare surface and (b) a possible reconstruction arrangement.

**Fig. 4.** (a) Schematic of graphene (blue) on the Cu(110) surface at a misorientation of 5$^o$, resulting in a periodic linear pattern at 5$^o$ to the [210] direction; (b) STM image of an area highlighting the odd-even transition, as indicated by arrows, (c) schematic of how this arises.

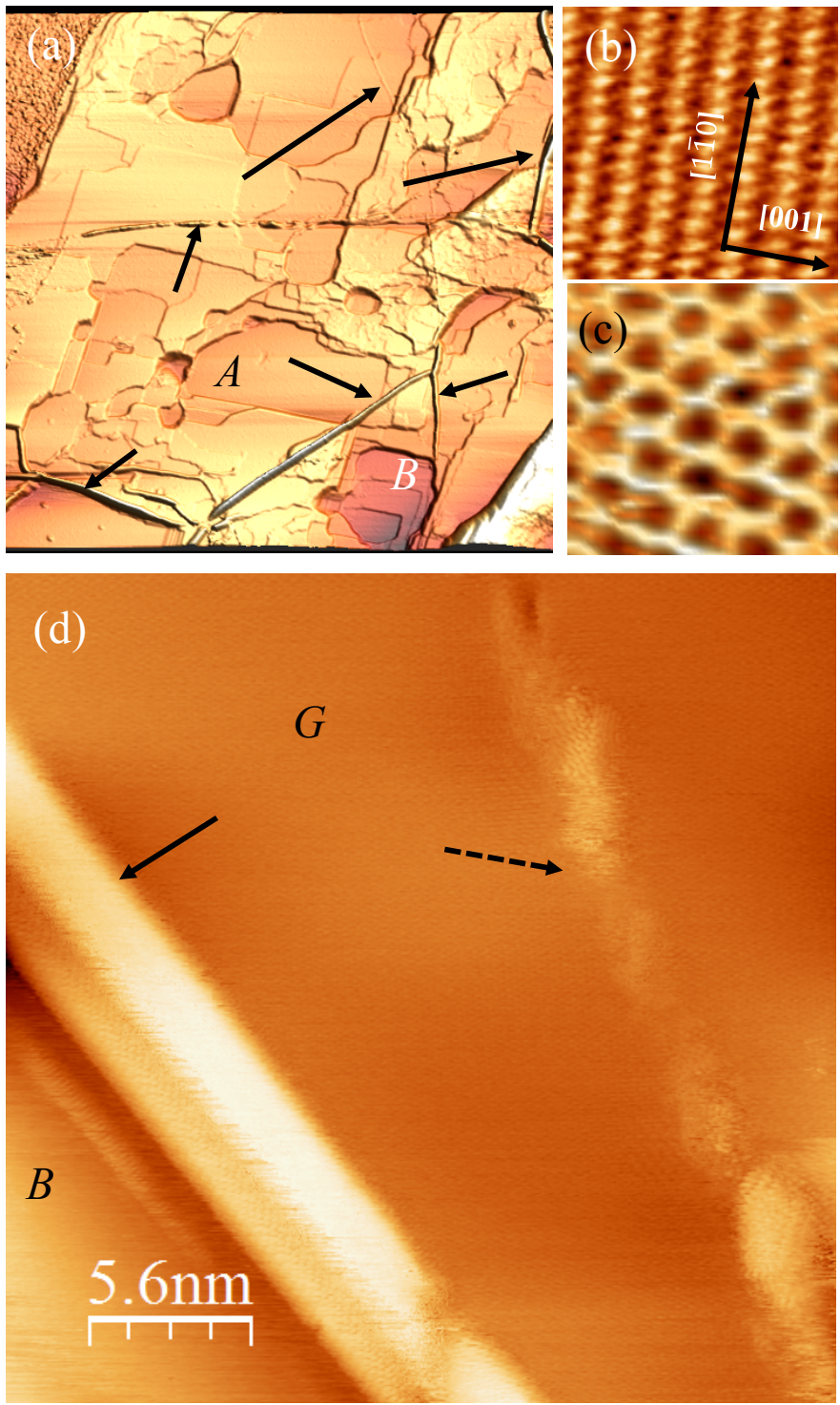

FIGURE 1. DURKAN

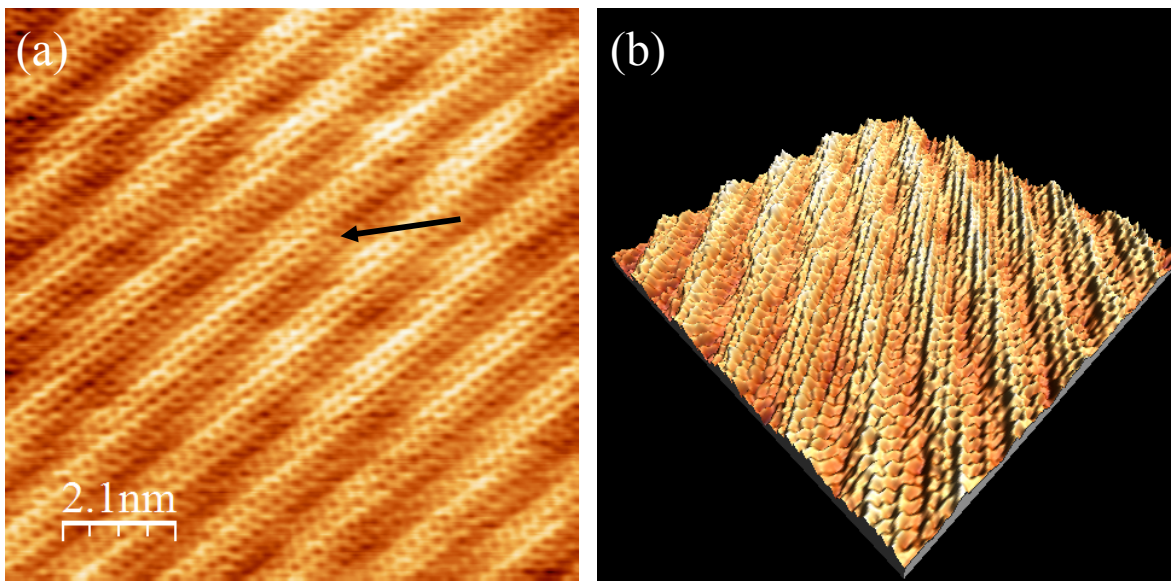

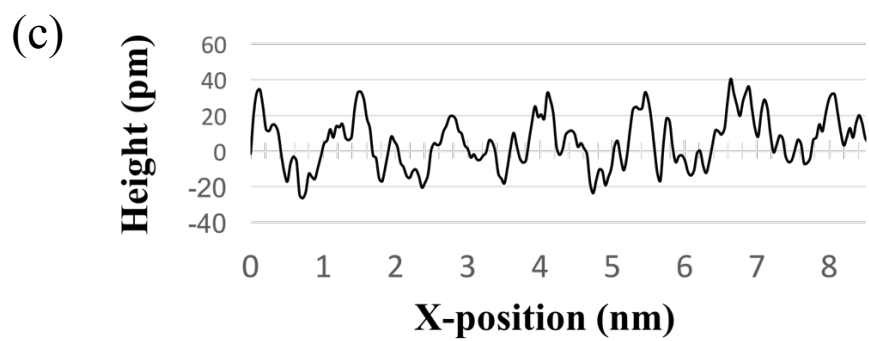

FIGURE 2. DURKAN

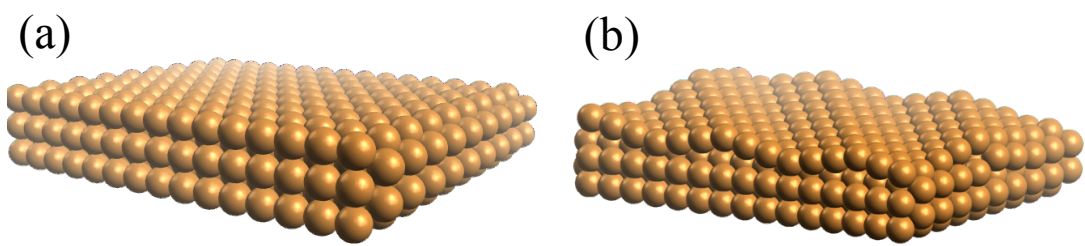

FIGURE 3. DURKAN

(a)

Direction of structures [210] 5°

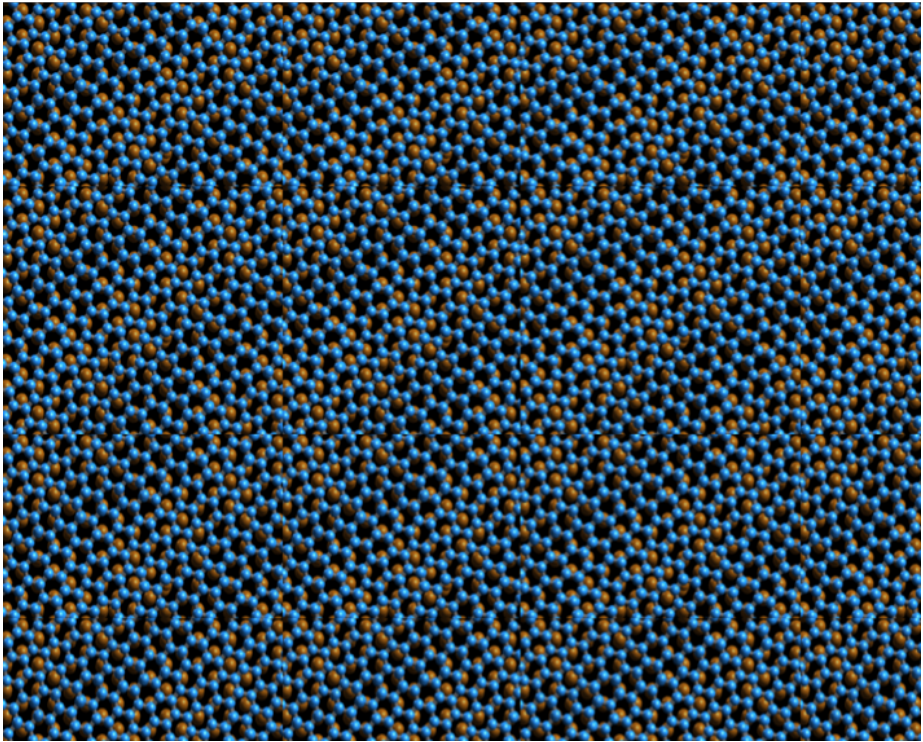

(b) 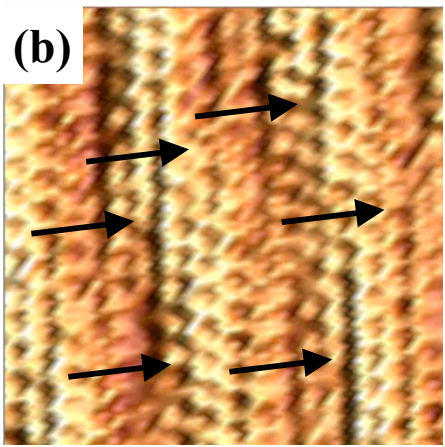

(c) 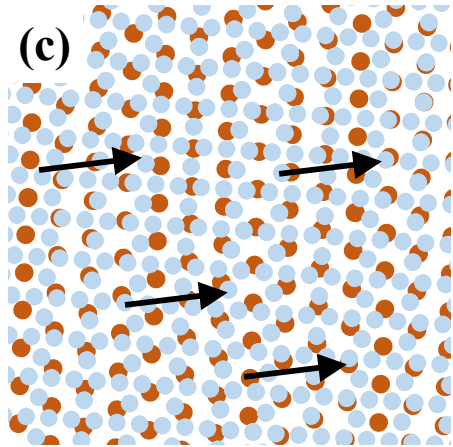

FIGURE 4. DURKAN

13